# Spectroscopic Analysis of the 2025 Eclipse of the Symbiotic Binary V1413 Aquilae

**David Boyd**
*West Challow Observatory, West Challow, Oxfordshire, UK; davidboyd@orion.me.uk*

**James R. Foster**
*800 Glenbrook Drive, Frazier Park, CA 93225; jrfcomet@sbcglobal.net*

**Forrest Sims**
*Desert Celestial Observatory, 5900 E Bell Street, Apache Junction, AZ 85119; forrest@simsaa.com*

**David Cejudo**
*Dark Sky Hosting Observatory (La Palma), Camino de las Canteras, 42; El Berrueco, Spain; davcejudo@hotmail.com*

**Robin Leadbeater**
*Three Hills Observatory, The Birches, Wigton, Cumbria, CA7 1JF, UK; robin@threehillsobservatory.co.uk*

**Pavol A. Dubovsky**
*Vihorlat Observatory, Kolonica 219, Kolonica, 06761, Slovakia; var@kozmos.sk*

**Xavier Dupont**
*4 rue de l'Orée du Bois, 37390 Saint-Roch, France; xdupont@wanadoo.fr*

**Keith Shank**
*Oaks & Stars Observatory, 304 S. Rolling Oaks Drive, Goldthwaite, TX 76844; kasism@verizon.net*

**David Trowbridge**
*Tinyblue Observatory, 3537 Gylany Way, Greenbank, WA; david.daiku.trowbridge@gmail.com*

**Jerry Hinnefeld**
*Indiana University South Bend, 1700 Mishawaka Avenue, South Bend, IN 46634; jhinnefe@iu.edu*

**Colin Eldridge**
*5 Booker Street, Attadale, Perth, Western Australia 6156; Colin-Eldridge@outlook.com*

**Joan Guarro i Fló**
*25354 Santa Maria de Montmagastrell Remote Observatory, Av, Catalunya, 38, Spain; jngrrfl@gmail.com*

**Kevin Gurney**
*Gladstone Road Observatory, Sheffield, UK; kevingurney42@gmail.com*

**Gary Walker**
*114 Cove Road, West Dennis, MA; bailyhill14@gmail.com*

**Franz-Josef Hambsch**
*Oude Bleken 12, Mol, B-2400, Belgium; hambsch@telenet.be*

**Abstract** We report on a coordinated campaign by amateur astronomers using a combination of high and low resolution spectroscopy to observe the 2025 eclipse of the symbiotic binary V1413 Aql shortly after a probable Z And-type symbiotic outburst. We measured the eclipse time of minimum and computed an updated eclipse ephemeris. Our spectra were calibrated in absolute flux and multi-Gaussian fits were used to resolve and quantify components of the Hα emission line. We measured the radial velocity curve of the white dwarf and determined the semi-amplitude as 25.7 km/s and the systemic radial velocity as 95.1 km/s. We found the spectral continuum was consistent with that of a late F or early G type giant star.

## 1. Introduction

Symbiotic stars are long period binaries comprising, typically, a cool red giant star and a hot white dwarf which orbits within a circumbinary nebula formed by the wind of the giant star (Merc 2025). UV radiation from the white dwarf ionizes a fraction of the red giant wind, producing nebular emission (Skopal 2003). Hydrogen-rich material accretes from the giant wind onto the surface of the white dwarf where it may undergo thermonuclear burning. A change in accretion can increase the burning rate creating an expanded white dwarf pseudo-photosphere. This expansion at roughly constant bolometric luminosity causes a reduction in its effective temperature. Radiation is redistributed from the UV to longer wavelengths, the system brightens visually, high ionization energy lines disappear, and the spectrum resembles a warm stellar continuum with low energy emission lines. We observe this as a Z And-type symbiotic outburst (Sokoloski *et al.* 2006). Between outbursts the system slowly returns to a quiescent state and radiation moves back into the UV. If the orbital inclination is sufficiently high, the hot component is eclipsed as it passes behind the giant. The symbiotic binary V1413 Aql is unusual in that outbursts occur relatively frequently and the system rarely returns to true quiescence before the next outburst begins. In V1413 Aql the giant star has spectral type M5III and eclipses occur every 434.1 days. For an analysis of recent observations of V1413 Aql see Tatarnikova *et al.* (2025).

We present analysis of high-cadence optical spectroscopy and multicolor $BVI_c$ photometry recorded during the 2025 eclipse of the symbiotic binary V1413 Aql. In section 2 we summarize our observations and data reduction. Section 3 describes our analysis and presents the results. Section 4 provides a summary.

## 2. Observations

The eclipse of V1413 Aql in 2025 was observed by the international team of amateur astronomers listed in Table 1 along with the low and medium/high resolution spectrographs they used and the spectra they contributed. We also used photometric observations of the 2025 eclipse of V1413 Aql reported to the AAVSO International Database (AID; Kloppenborg 2026a) by observers worldwide, including members of the team, in B, V, and $I_c$ standard filters as shown in Figure 1. Observations were coordinated through meetings of the online Spectroscopy Discussion Group (2026) organized by Forrest Sims.

Spectra were analyzed by the observer using similar procedures to those described in Boyd *et al.* (2025) (hereafter Paper 1) and submitted to one or more of the following spectroscopy databases: ARAS (ARAS Spectral Database; Teyssier 2019), AAVSO (AVSpec; Kloppenborg 2026b), and BAA Spectroscopy Database (Br. Astron. Assoc. 2026).

The first step in processing all spectra after downloading them from the databases was to read the FITS header, check for the presence of necessary keywords, add missing keywords required for downstream processing, introduce consistency where different keywords were required for the same item by different databases, and remedy deficiencies such as the omission of geographical location information required to calculate barycentric correction. The appropriate barycentric correction was then calculated using the routine pyasl.helcorr (PyAstronomy 2026) and applied. Spectra from all sources were combined, duplicates removed, and as a final step, the wavelength calibration of each high resolution spectrum was corrected using eight telluric lines close to Hα. Analysis of the 150 spectra obtained was carried out using a range of publicly available software and several custom PYTHON codes making use of Astropy resources (Astropy Collaboration *et al.* 2018).

## 3. Analysis and results

### 3.1. Eclipse time of minimum and updated ephemeris

The AAVSO light curve between 2023 and 2025 (Figure 2) shows V1413 Aql experienced three eclipses while gradually brightening towards a V magnitude of ~10.4 during solar conjunction in early 2025. Analysis of the 2023 eclipse was reported by Poyner (2023) and of the 2024 eclipse in Paper 1.

Photometry in B, V, and $I_c$ from the AAVSO AID was analyzed to obtain a time of minimum in each filter by quadratic analysis of the eclipse profile and these were combined to obtain the mean mid-eclipse time for the 2025 eclipse as JD 2460981.35 ± 0.31 (2025 November 1.85).

Publications on V1413 Aql have for many years referred to the ephemeris of eclipses of the hot component in Munari (1992), JD = 2446650(± 15) + 434.1(± 0.2) E, recently refined by Tatarnikova *et al.* (2025) to JD = 2448819.50 (± 1.5) + 434.45 (± 0.27) E.

We analyzed observations of previous eclipses in the AAVSO AID obtained visually or with V or Clear filters to obtain the mid-eclipse times of a further 16 eclipses listed in Table 2. These were used to calculate the following updated linear ephemeris:

$$JD = 2446654.19\ (\pm\ 1.44) + 434.154\ (\pm\ 0.076)\ E.$$

Figure 3 shows the O–C difference between the observed mid-eclipse times and those calculated using this new linear ephemeris.

### 3.2. Eclipse analysis

After considering together the profiles of the eclipses in 2023, 2024, and 2025 in Figure 2, we revise the conclusion in Paper 1 that the 2024 eclipse was total and now consider all three eclipses to have been partial. We measured the times of 1st and 4th contact of the 2025 eclipse by intersecting a linear interpolation of the V band flux profile across the eclipse with linearly extrapolated ingress and egress slopes of the eclipse. This gave an eclipse duration of 72 days. The 2023, 2024, and 2025 eclipses had depths of 2.0, 1.9, and 1.3 magnitudes, respectively, as increasing activity caused expansion of the hot component pseudo-photosphere in the lead-up to a Z And-type outburst which probably occurred early in 2025.

By identifying closely concurrent photometric observations in multiple filters, we calculated the B–V and $V–I_c$ color indices shown in Figure 4. Prior to eclipse B–V = 0.86 ± 0.02. During eclipse it became slightly bluer at B–V = 0.80 ± 0.04, similar to behavior reported during the 2024 eclipse in Paper 1. Prior to the eclipse $V–I_c$ = 1.27 ± 0.03. In all cases these uncertainty values are standard deviations. During the eclipse, between phases –0.037 and 0.037, a period of 32 days, $V–I_c$ rose rapidly to a peak of $V–I_c$ = 1.5. This period is shown shaded in Figure 4. A similar rapid and brief reddening during the 2024 eclipse was reported in Paper 1 and during the 2023 eclipse in Tatarnikova *et al.* (2025). This steep-sided reddening peak could be attributed to eclipse by the red giant star of a small relatively blue object.

Table 1. Observers contributing spectra with equipment used.

| *Observer* | *Equipment* | *Resolving Power* | *Number of Spectra* |
|---|---|---|---|
| Boyd | C11 + LISA | 1100 | 3 |
| Boyd | C11 + StarEx2400 | 10600 | 21 |
| Cejudo | 0.35m Ritchey-Chretien + UVEX600 | 1750 | 21 |
| Dubovsky | C14 + NOU_T echelle | 10880 | 5 |
| Dupont | DK 350 + StarEx300 | 730 | 1 |
| Dupont | DK 350 + StarEx2400 | 15000 | 2 |
| Eldridge | CDK1000 + eShel | 10000 | 1 |
| Foster | 0.33m Cassegrain + ALPY600 | 480 | 13 |
| Foster | CDK17 + ALPY600 | 480 | 8 |
| Foster | CDK17 + eShel | 11000 | 12 |
| Guaro Flo | SCT16 + NOU_T echelle | 9000 | 1 |
| Gurney | C11 + ALPY600 | 600 | 1 |
| Hinnefeld | SCT16 + ALPY600 | 530 | 2 |
| Leadbeater | C11 + ALPY600 | 620 | 6 |
| Leadbeater | C11 + LHIRES1200 | 5050 | 7 |
| Leadbeater | C11 + LHIRES2400 | 14800 | 7 |
| Shank | C14 + LISA | 1000 | 3 |
| Sims | CDK20 + LISA | 1000 | 6 |
| Sims | CDK20 + NOU_T2 echelle | 9000 | 21 |
| Teyssier | SCT16 + NOU_T echelle | 9000 | 6 |
| Trowbridge | Mewlon250 + DADOS200 | 910 | 3 |

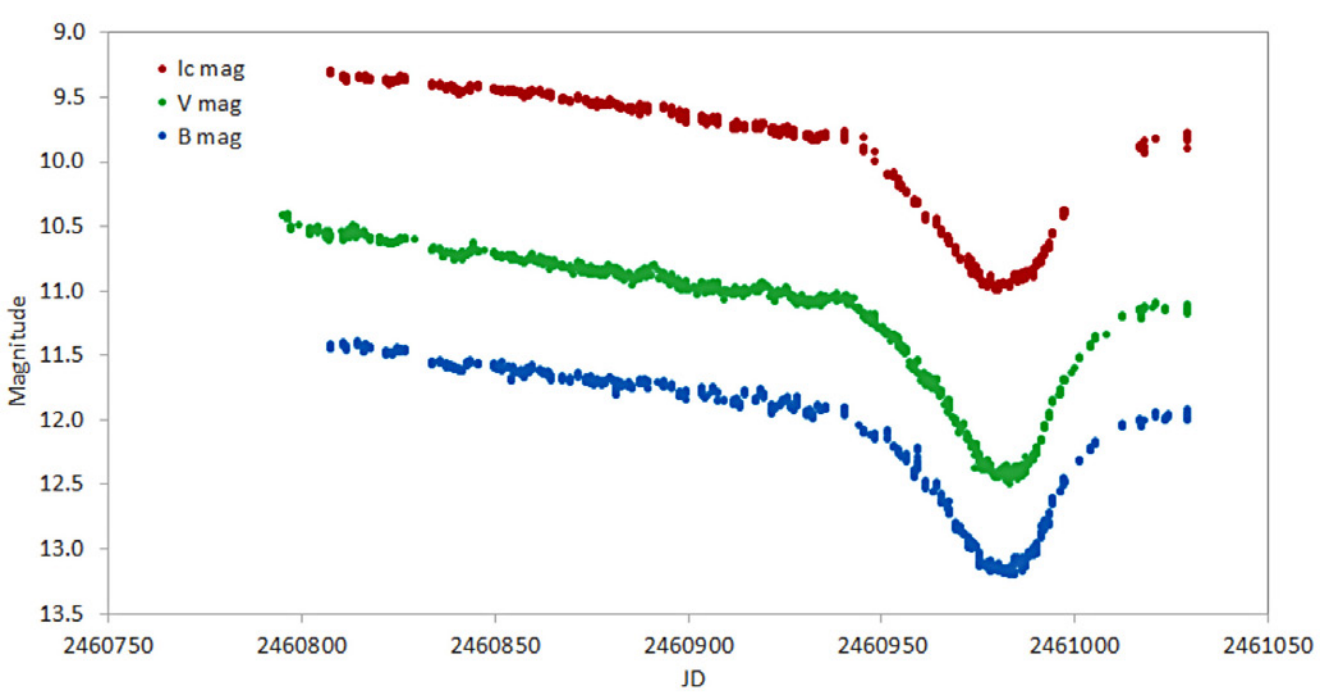

Figure 1. Photometric observations of the 2025 eclipse of V1413 Aql reported to the AAVSO AID by many observers in B, V, and $I_c$ standard filters.

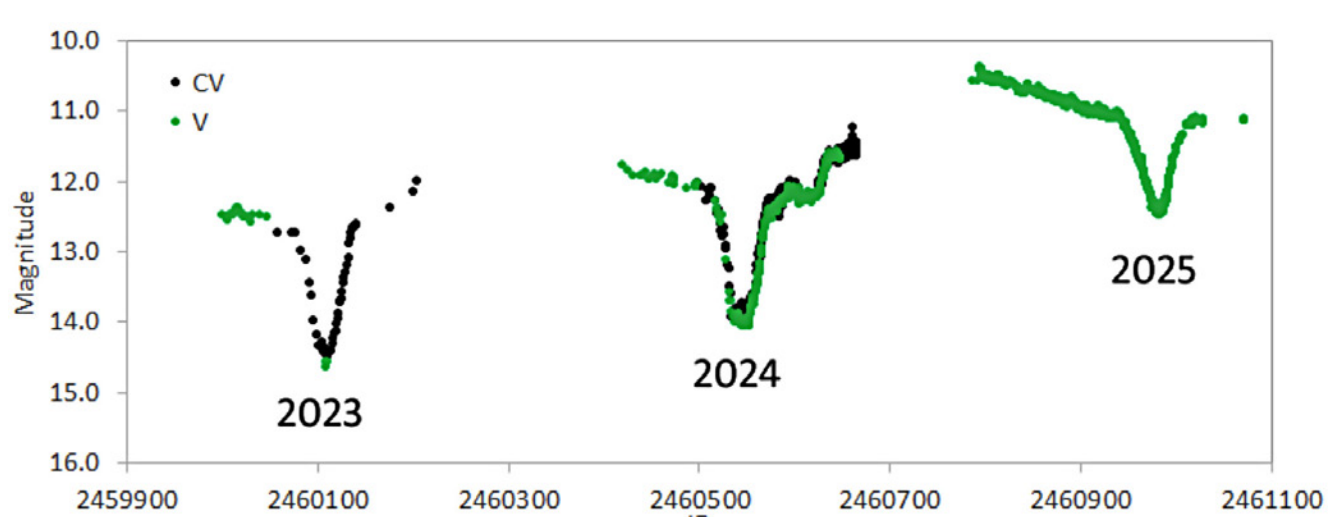

Figure 2. AAVSO light curve of V1413 Aql between 2023 and 2025 recorded with either a V filter (V) or a clear filter using V-band magnitudes (CV).

Table 2. Observed mid-eclipse times for 17 eclipses.

| *Year* | *Eclipse Number* | *Observed Mid-eclipse Times (JD)* |
|---|---|---|
| 1986 | 0 | 2446650.00 |
| 1987 | 1 | 2447087.00 |
| 1991 | 5 | 2448822.00 |
| 1997 | 9 | 2450564.06 |
| 1998 | 10 | 2450999.90 |
| 1999 | 11 | 2451429.43 |
| 2000 | 12 | 2451869.00 |
| 2004 | 15 | 2453164.15 |
| 2005 | 16 | 2453599.58 |
| 2006 | 17 | 2454041.02 |
| 2010 | 20 | 2455339.00 |
| 2011 | 21 | 2455769.00 |
| 2011 | 21 | 2455770.88 |
| 2012 | 22 | 2456205.04 |
| 2023 | 31 | 2460110.45 |
| 2024 | 32 | 2460545.80 |
| 2025 | 33 | 2460981.35 |

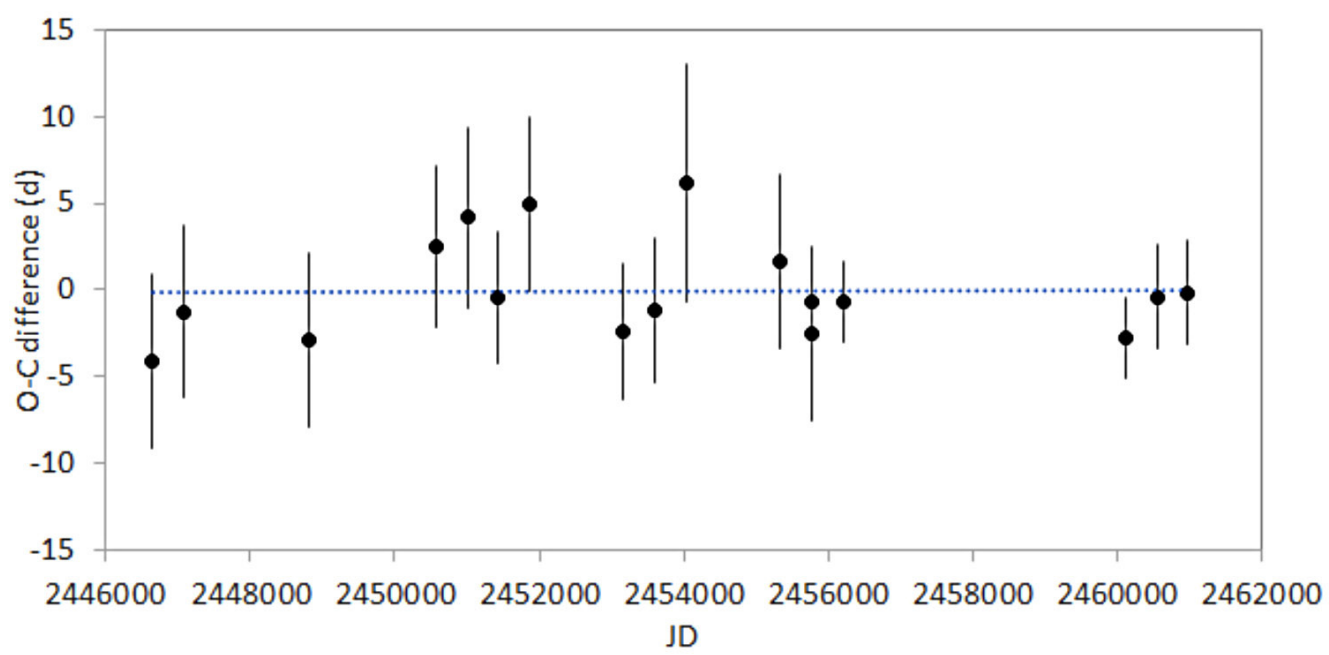

Figure 3. O–C difference between observed mid-eclipse times and those calculated from the updated linear ephemeris: JD = 2446654.19 (± 1.44) + 434.154 (± 0.076) E.

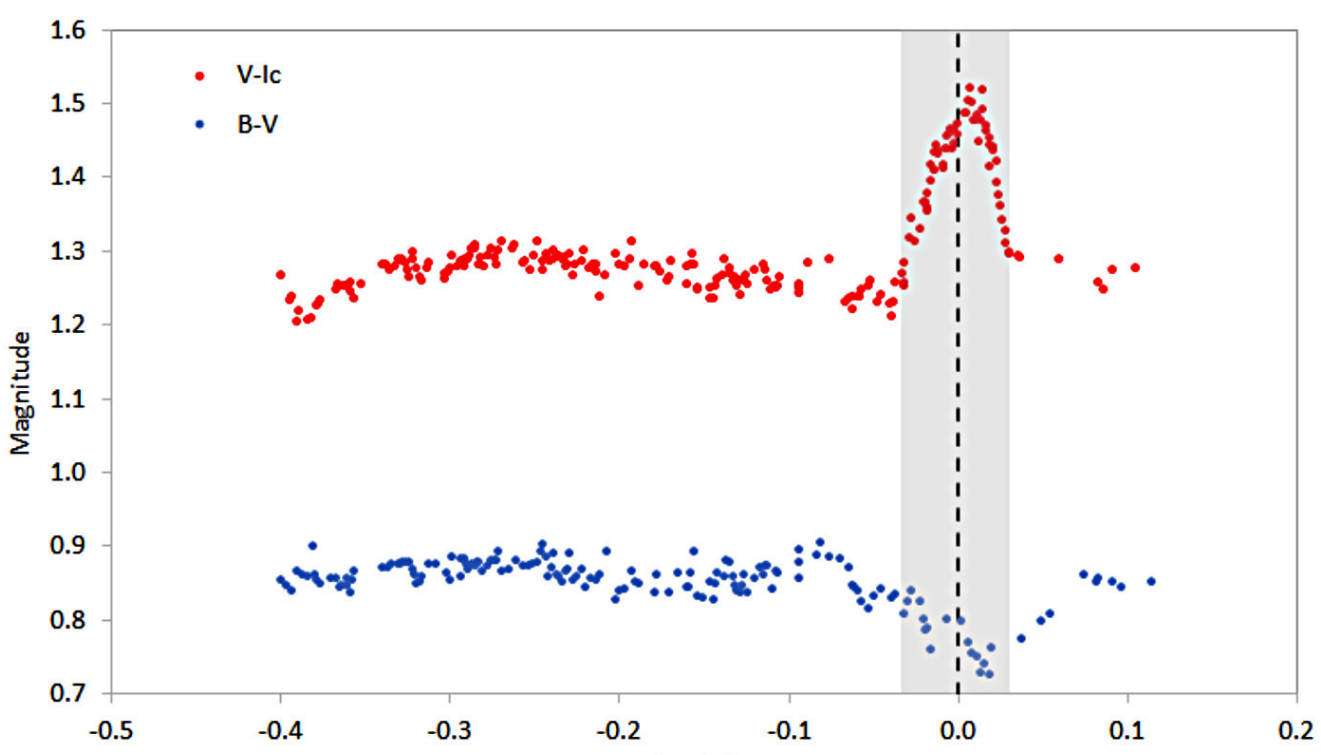

Figure 4. Variation of B–V and V–$I_c$ color indices with orbital phase. The photometric eclipse is between orbital phases –0.08 and +0.08. The rapid increase in V–$I_c$ between phases –0.037 and +0.037 is shown shaded.

### 3.3. Calibrating spectra

To calibrate spectra in absolute flux we needed to know the V magnitude at the time of each spectrum. Some team members measured the V magnitude of V1413 Aql concurrently with recording its spectrum and reported this magnitude to AAVSO AID. We downloaded all V band data from the AAVSO AID, removing obvious outliers, calculating daily averages and five-day running averages to obtain a smoothed V band light curve. Attempts to find flickering in V1413 Aql in the past have failed (Paper 1 and Tatarnikova *et al.* 2025). We therefore consider this smoothed light curve to be an accurate representation of its time-varying V magnitude and estimate its uncertainty by comparison with the original V-band data as less than ± 0.05 magnitude. This was then interpolated to obtain a concurrent V magnitude at the time of each spectrum and this was added as a keyword to the FITS header for use in absolute flux calibration.

All our low-resolution spectra from 2024 and 2025 which covered both the complete V filter bandpass and the region around Hα were calibrated in absolute flux units of erg/cm$^2$/s/Å using the method described in Boyd (2020). The absolute flux of the continuum at Hα could then be found for each of these spectra. After sigma-clipping to remove outliers caused by poor instrument response correction, Figure 5a shows there was a strong correlation over 2 years and 3.5 magnitudes between the logarithm of continuum flux at Hα and concurrent V magnitude. We used this relationship, and the concurrent V magnitude for each spectrum, to establish the absolute flux at Hα for each spectrum which included the region around Hα, including those at higher resolution which did not include the V passband. From the scatter in Figure 5a, we estimate the uncertainty from determining absolute line flux in this way to be less than ± 5%. The largest contributors to this uncertainty are in calculating the interpolated V magnitude and in the accuracy of response correction in the low resolution spectra used to construct the above relationship. A similar approach was used to establish the relationship between the logarithm of continuum flux at Hβ and concurrent V magnitude in Figure 5b. This was used to establish the absolute flux at Hβ for spectra in the region around Hβ. Establishing a separate relationship at Hβ was necessary to obviate problems with relative flux calibration of the continuum of some echelle spectra and this calibration was used to analyze emission lines blue-ward of 5500 Å. Establishing the absolute flux of each spectrum enables quantitative comparison of line fluxes obtained with different instruments at different resolutions and at different times.

Gaia DR3 (Bailer-Jones *et al.* 2021) gives the distance of V1413 Aql as 5.6 (–1.0 +1.4) kpc. V1413 Aql experiences interstellar extinction with a color excess generally accepted to be E(B–V) = 0.5 ± 0.05 (Esipov *et al.* 2000; Tatarnikova *et al.* 2025). We used this color excess with formulae in Cardelli *et al.* (1989), assuming R(V) = 3.1, to calculate A(λ)/A(V) and applied this wavelength-dependent dereddening to all our absolute flux calibrated spectra.

### 3.4. Analysing low resolution spectra

The light curve in Figure 2 shows V1413 Aql probably peaked around V magnitude ~10.4 early in 2025. A composite plot of our dereddened low resolution spectra obtained prior to and during the 2025 eclipse is shown in Figure 6. In 2024 TiO molecular bands of the M-type giant star were visible in spectra at eclipse minimum (Paper 1). In 2025, these molecular bands were overwhelmed, largely by continuum light from un-eclipsed nebular emission which contributes to the minimum of the 2025 eclipse being 2 magnitudes brighter than in 2023 and to the absence of molecular bands at eclipse minimum in Figure 6. Strong nebular continuum throughout the optical was also seen in AX Per (Skopal *et al.* 2011, Figure 8).

Recombination lines of H I and Fe II, generated by ionising radiation with photon energies > 25 eV, were prominent and remained visible throughout the eclipse, indicating they were formed in the un-eclipsed region of the nebula by radiation from the hot component. He I lines at 5876 Å and 6678 Å were weakly but clearly detectable. Forbidden lines of [O III] and [N II] which had been seen in 2024 were not detectable, suggesting lower photon energies than in 2024. There were no high ionization lines such as He II. This behavior is typical of a Z And-type symbiotic outburst. Profiles of the Na I D1 and D2 lines were complex, changing from absorption to emission at eclipse minimum. The diffuse interstellar band at 6283 Å was conspicuous.

We estimated temperatures corresponding to these spectra by comparing the continuum of our dereddened low resolution spectra in the range 4750 to 6800 Å to Planck black body spectra. Figure 7 shows examples of our spectra recorded prior to the eclipse and at eclipse minimum together with Planck spectra for 6200 K and 5600 K, respectively. Note however that the continuum of these spectra included contributions from nebular emission as well as the hot component pseudo-photosphere so these were not reliable estimates of the hot component temperature. These temperatures correspond to a late F or early G type giant star and are similar to the temperature of the flared disk whose rim formed the warm pseudo-photosphere in AX Per (Skopal *et al.* 2011).

Absolute flux in erg/cm$^2$/s of emission lines in our low resolution spectra was found by numerically integrating the line profile above an interpolated linear continuum. The uncertainty in line flux was estimated from the uncertainties in determining the spectral profile of the line, in estimating the flux of the interpolated continuum under the line, and in the relationship between Hα continuum flux and V magnitude. Figure 8 shows how the integrated flux in Hα, Hβ, and Hγ in 2025 varied with orbital phase while Figure 9 shows the same for six low ionization Fe II lines. Flux in the Balmer lines reduced by ~40% during the eclipse with a smaller reduction in flux in the Fe II lines. This behavior is broadly similar to that seen in

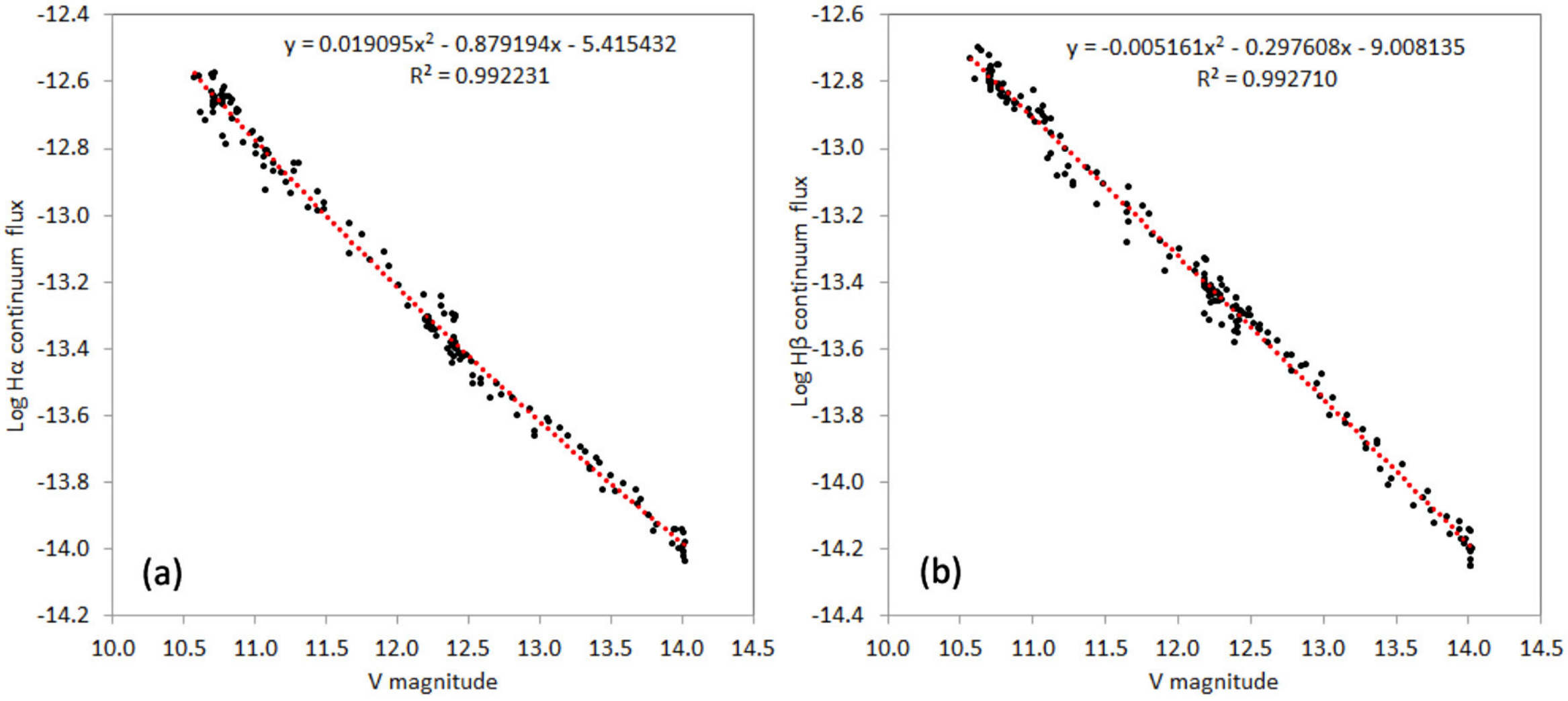

Figure 5. Correlation of concurrent V magnitude with logarithm of continuum flux at Hα (a) and Hβ (b).

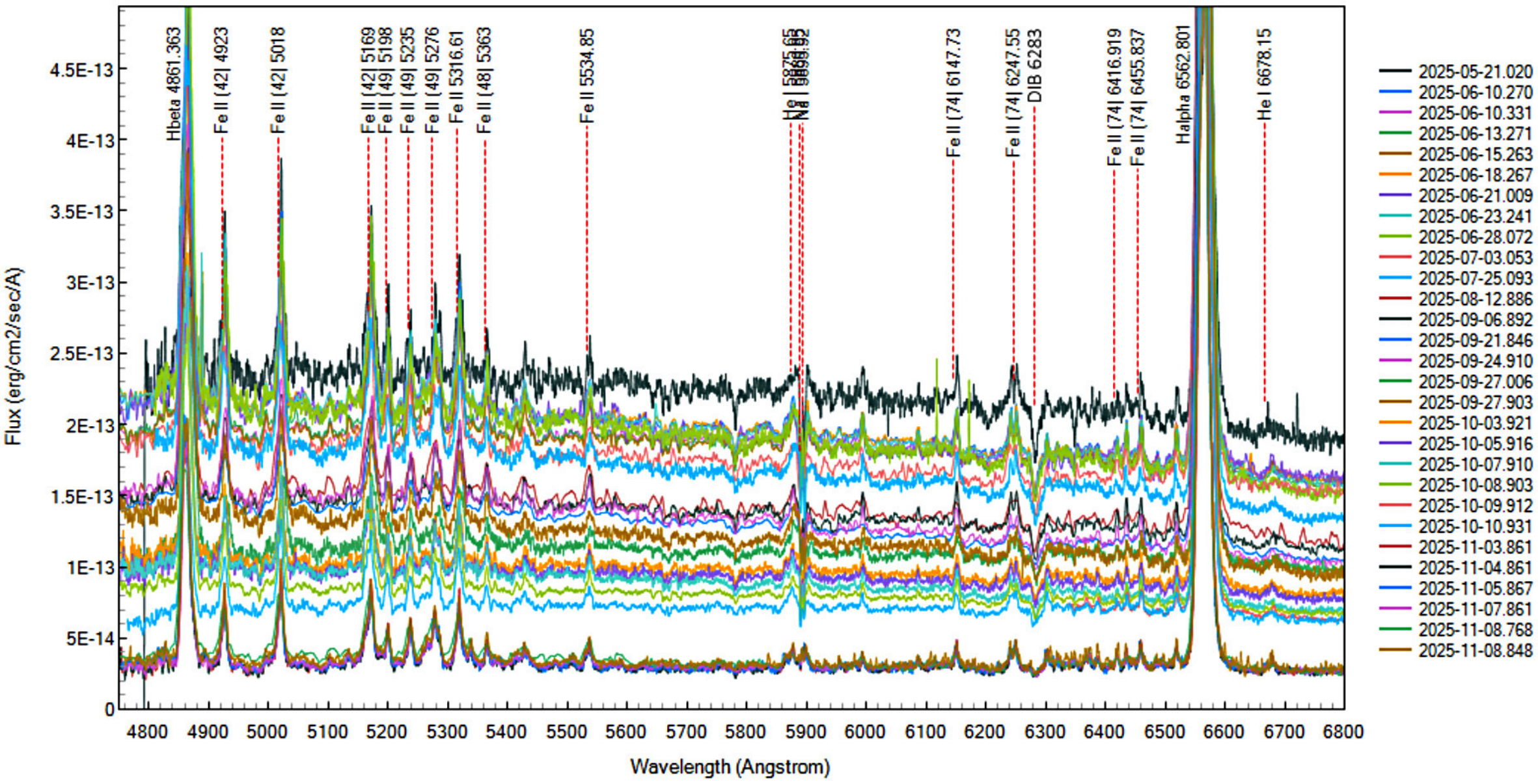

Figure 6. Composite plot of our dereddened low resolution spectra obtained prior to and during the eclipse in 2025 with prominent features identified. The lowest flux spectra were recorded at eclipse minimum and show no evidence of TiO molecular bands.

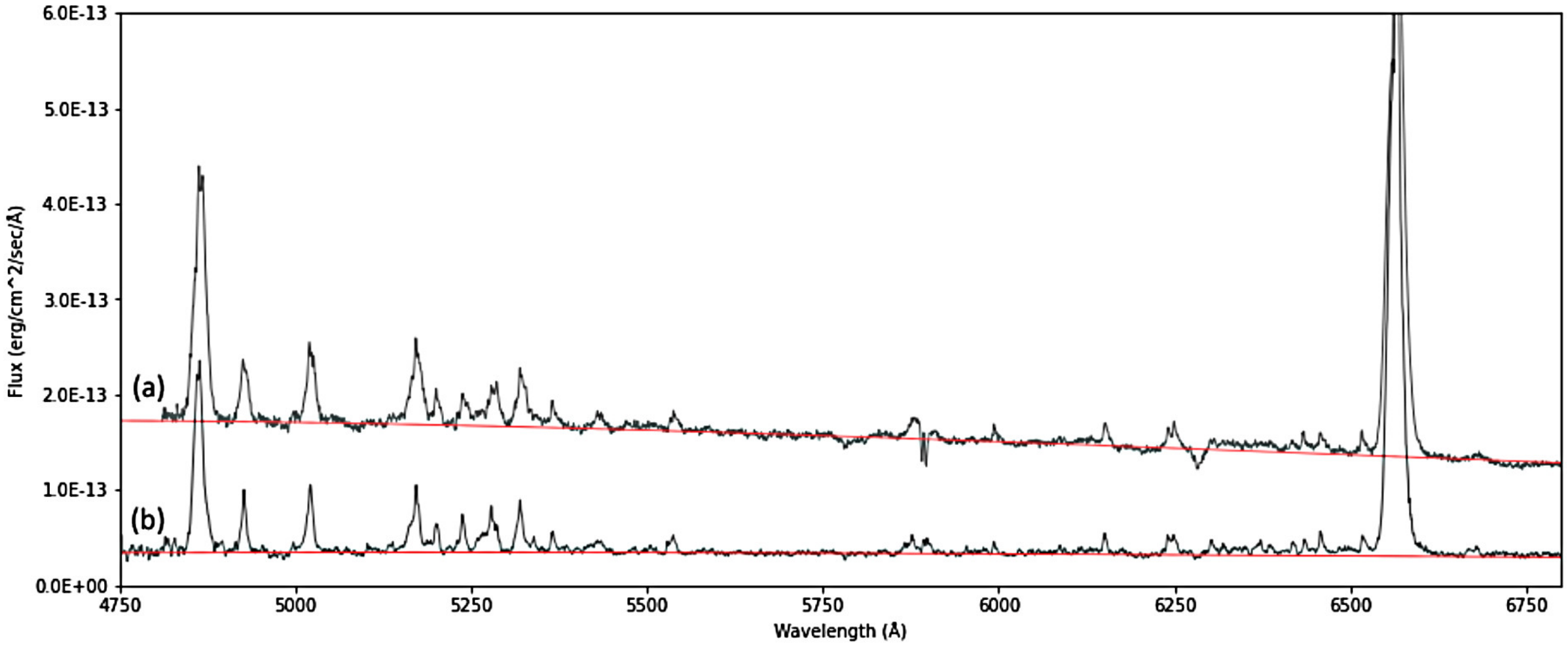

Figure 7. Spectra recorded prior to eclipse (a) and at eclipse minimum (b) with (in red) Planck black body spectra for 6200 K and 5600 K respectively.

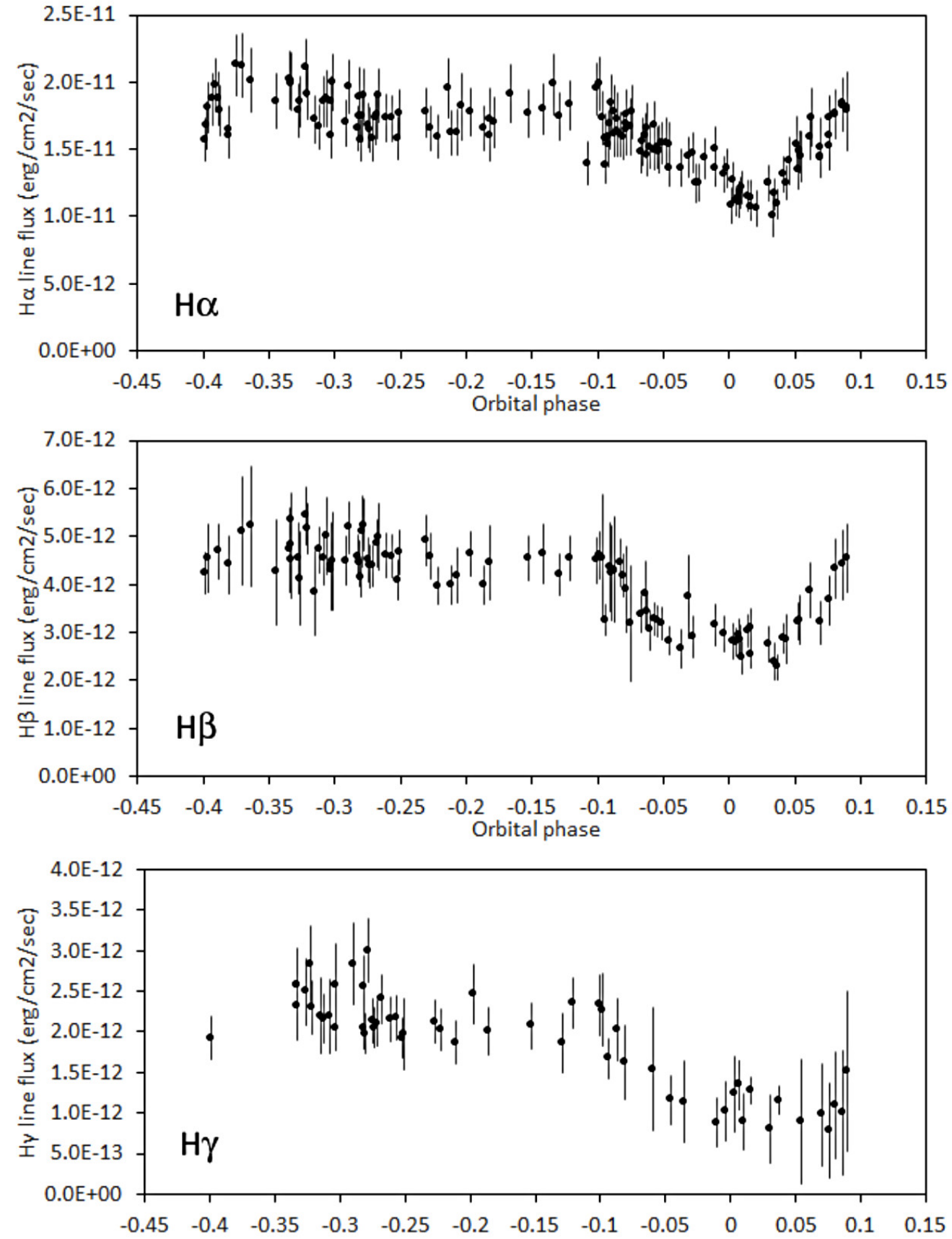

Figure 8. Variation of absolute flux in Hα, Hβ, and Hγ emission lines with orbital phase in 2025.

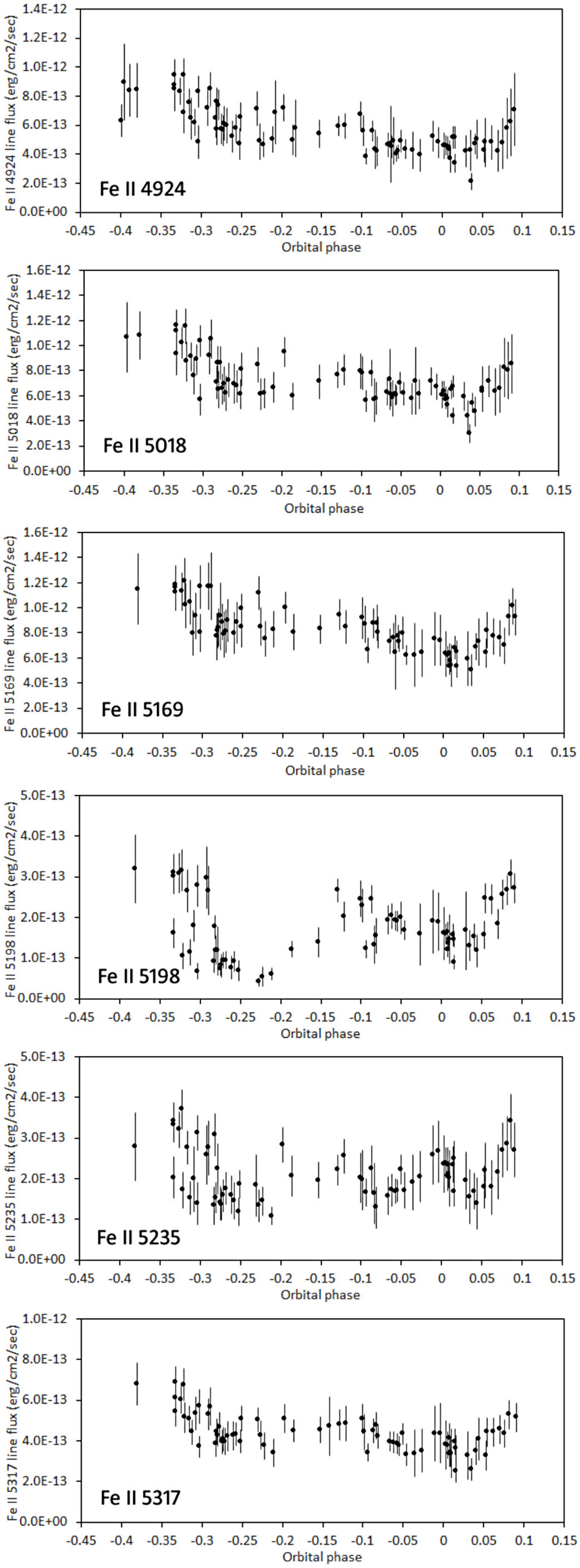

Figure 9. Variation of absolute flux in Fe II emission lines with orbital phase in 2025.

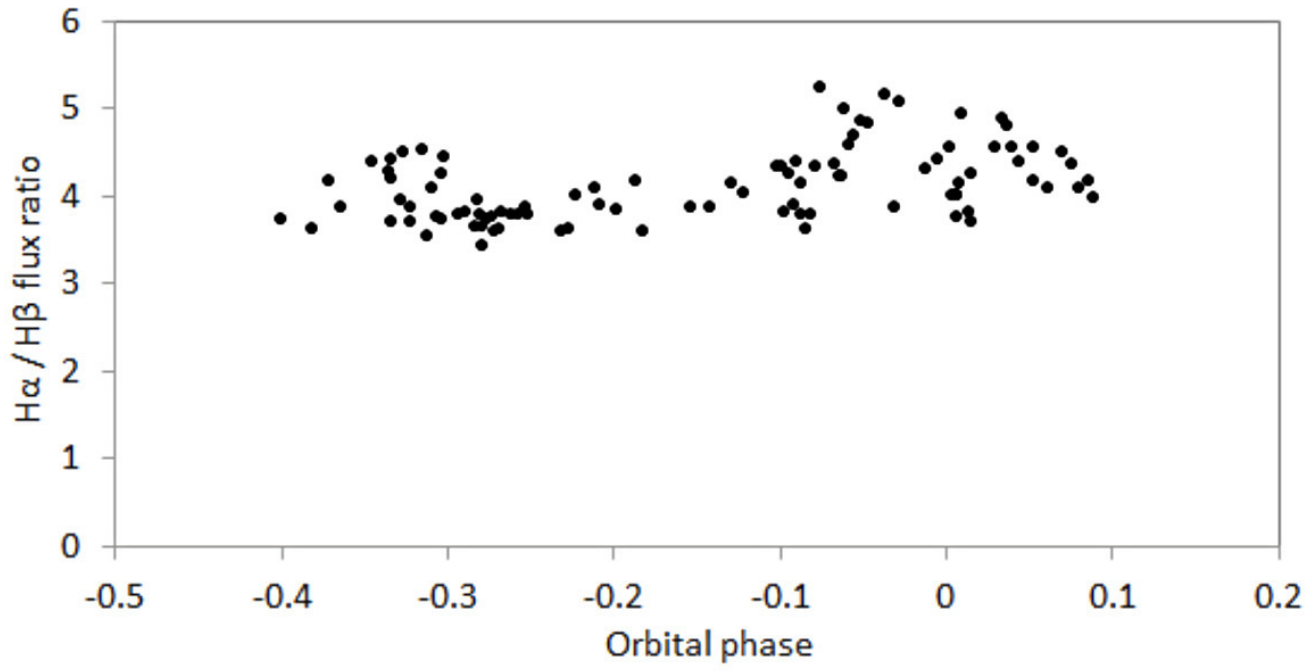

Figure 10. Variation of Hα/Hβ emission line flux ratio with orbital phase in 2025.

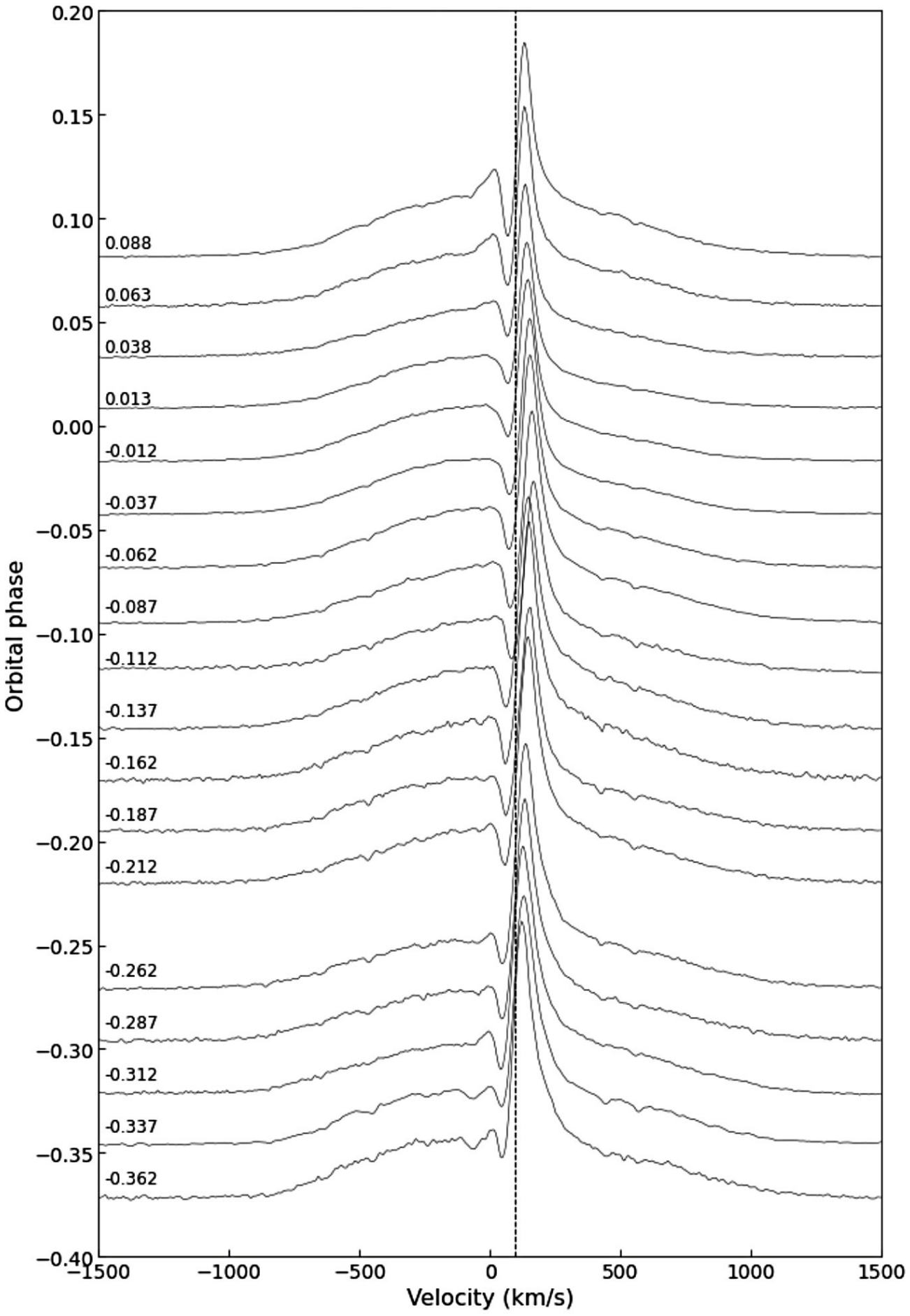

Figure 11. Average spectral flux profiles of the Hα emission line in bins of equal width in phase with the derived systemic velocity shown as a dotted line.

2024. Integrated flux in the H I lines was ~2.5 times larger in 2025, as expected with the star one magnitude brighter, while flux in the Fe II lines was ~5 times larger. Some of the Fe II lines showed noticeable decreases in flux in the period prior to the eclipse while some showed small increases in flux in the center of the eclipse. The mean and standard deviation of the Hα/Hβ emission line flux ratio was 3.90 ± 0.28 outside eclipse and 4.32 ± 0.41 during eclipse (Figure 10).

### 3.5. Analysing higher resolution spectra

Emission lines in our higher resolution spectra have narrow emission and absorption components and a broad emission base. The blue-shifted absorption component appears consistent with a P Cygni profile. We aggregated all our high resolution spectra of the Hα line in bins of equal width in phase, calculated an average spectral flux profile for each bin, and plotted these in Figure 11.

By empirically modelling these emission lines with three Gaussian profiles, one for the broad base and one each for the narrow emission and absorption components, we are able to quantify these components. Simultaneously fitting all three components to the observed profile of the line gives the peak flux in $erg/cm_2/s/Å$, the wavelength at the peak, and the full width at half maximum (FWHM), both in Å, for each component. The displacement of the peak wavelength relative to the rest wavelength of the line, and the FWHM, can be converted into radial velocities in km/s. Figure 12 shows examples of Gaussian fits to Hα, Hβ, and Fe II 5018 Å emission lines. To check the consistency and integrity of this analysis, we compared the combined flux of the three Gaussians to the numerically integrated flux. These were within 1% for Hα and within 1.5% for Hβ and Fe II 5018 Å.

The profile of these lines changed markedly between 2024 and 2025. In 2025 the base of the line was much broader and bell-shaped. Figure 13 compares profiles of Hα emission lines at orbital phase –0.06 in 2024 and 2025 together with the three component Gaussian fits to the observed line profiles in each year. In 2024 the FWHM of the broad base component of the line was 600 km/s at that phase while after the probable symbiotic outburst in 2025 this had increased to 980 km/s.

### 3.6. Orbital parameters and radial velocities

Adopting component parameters used in Paper 1 and with an orbital period of 434.1 days, Kepler's Third Law gave a semi-major axis of 1.36 AU and a projected orbital velocity for the white dwarf of 21 km/s. From the redshift of multiple emission lines in low resolution spectra, Paper 1 estimated a systemic radial velocity for V1413 Aql of 92 km/s.

Figure 14 shows how the radial velocity of the narrow emission component in our Hα high resolution spectra varies with orbital phase. This may be interpreted as part of a circular orbit where the data only covers half the orbit. A sine fit to these data gives the radial velocity curve in Figure 15a with mean radial velocity 95.07 ± 0.32 km/s and semi-amplitude 25.67 ± 0.49 km/s. Figure 15b shows this radial velocity curve after subtraction of the mean radial velocity. Zero radial velocity occurs close to orbital phase zero, superior conjunction of the red giant. Maximum negative radial velocity of 25.7 km/s occurs at phase 0.25 and maximum positive radial velocity at phase 0.75. This is consistent with the orbital behavior expected of the white dwarf in a circular orbit. Taking the source of this narrow emission as co-located with the white dwarf, and noting the similarity of the mean radial velocity of this orbit to the estimated systemic binary radial velocity found in 2024, we conclude that Figure 15b also represents the radial velocity curve of the white dwarf and that 95 km/s is the true systemic radial velocity of V1413 Aql.

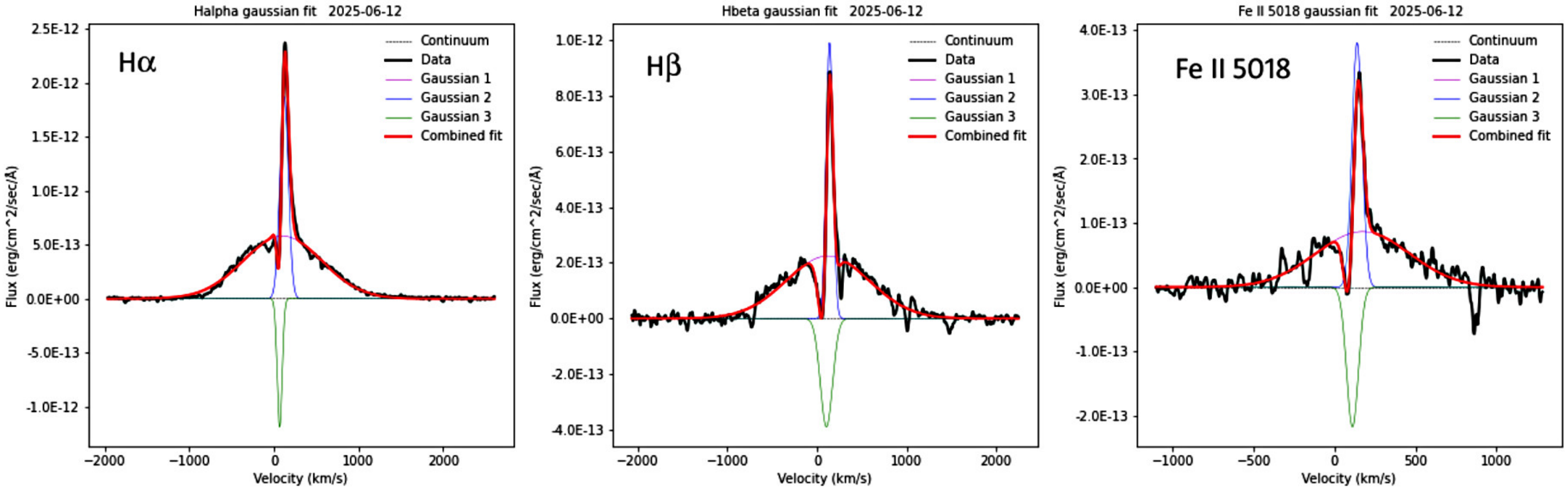

Figure 12. Hα, Hβ, and Fe II 5018 emission lines fitted with three Gaussian components.

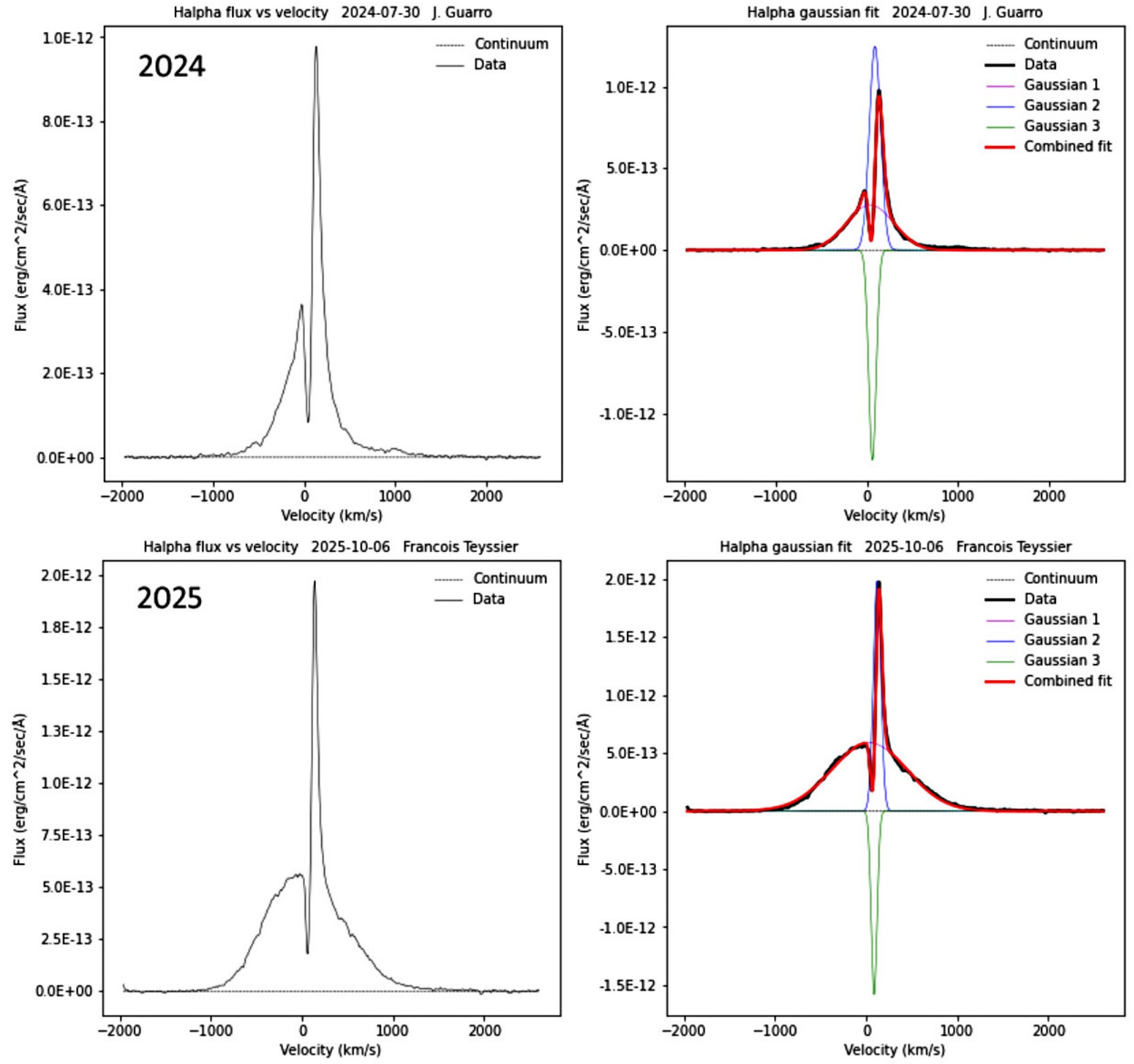

Figure 13. Comparison of Hα emission lines, both at orbital phase –0.06, in high resolution spectra in 2024 and 2025 plus fits with three Gaussian components in each case to the observed line profile.

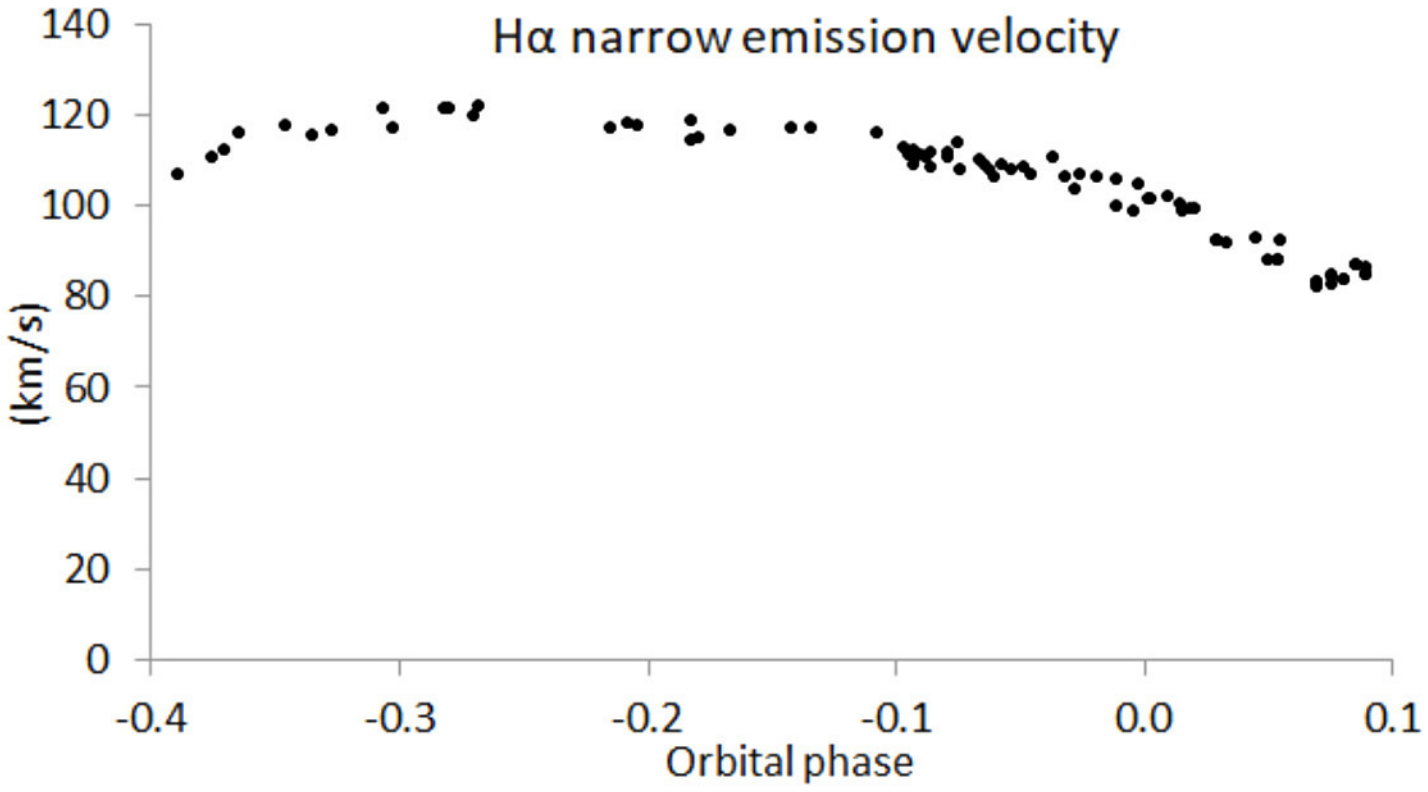

Figure 14. Variation of radial velocity of the narrow emission component in our Hα high resolution spectra with orbital phase.

From the dynamics of this orbit, the projected orbital velocity of the white dwarf will be ~26 km/s and assuming an orbital inclination of 68° from Paper 1 its velocity in the orbital plane will be ~28 km/s. To achieve this orbital velocity with a 0.6 $M_\odot$ white dwarf, Kepler's Third Law requires the mass of the red giant to be ~1.8 $M_\odot$. This is consistent with the mass of red giants in other symbiotic binaries (Fekel *et al.* 2000; Mikołajewska 2003; Zamanov *et al.* 2007). The semi-major axis would then be ~323 $R_\odot$ (~1.50 AU) and the radii of the white dwarf and red giant orbits ~242 $R_\odot$ and ~81 $R_\odot$, respectively. From Table 7 in van Belle *et al.* (1999), a plausible radius of the M5 giant star is 140 $R_\odot$ (0.65 AU). The 2025 photometric eclipse lasted 72 days and, with the above orbital parameters, the radius of the extended hot component being eclipsed can be calculated as ~77 $R_\odot$ (~0.36 AU).

From Eggleton (1983) we can calculate Roche lobe radii of the red giant and white dwarf with masses of 1.8 $M_\odot$ and 0.6 $M_\odot$ as 0.72 AU and 0.43 AU, respectively, so both the red giant and the extended hot component would be within their respective Roche lobes.

### 3.7. Analysing Hα emission

After subtracting the systemic radial velocity, Figure 16 shows how radial velocities and FWHM of the broad emission (a, b), narrow emission (c, d), and narrow absorption (e, f) components of the Hα line in our high resolution spectra vary with orbital phase.

We interpret these observations in the context of the analysis of symbiotic binaries in Skopal (2023). During a Z And-type symbiotic outburst, the enlarged white dwarf pseudo-photosphere forms a disk with a flared rim. At the orbital inclination of V1413 Aql, the broad rim of the disk occults the white dwarf at its center so we only see the warm stellar continuum of the pseudo-photosphere. UV radiation from the white dwarf emerges out of the plane of the disk and ionizes the surrounding nebula. For a pictorial representation, see Figure 2 in Skopal (2023).

Following an outburst, the optically thick expanding pseudo-photosphere generates a dense stellar wind (Sokoloski *et al.* 2006; Bisikalo *et al.* 2006). Broad Hα emission lines with FWHM ~1100 km/s are formed in the densest region of this wind where the velocity dispersion is large (Figure 16a). During the eclipse, the wind is partially obscured, and the observed width of the line reduces. Figure 17a compares the profiles of Hα lines recorded before and during the eclipse. During the eclipse, red-shifted emission from the far side of the disk is differentially obscured resulting in the balance of emission being displaced towards the blue. As a result the radial velocity of the broad component of the line is displaced to ~–120 km/s (Figure 16b). Figure 17b shows that this is more extreme in the case of Hβ emission which is displaced to ~–200 km/s.

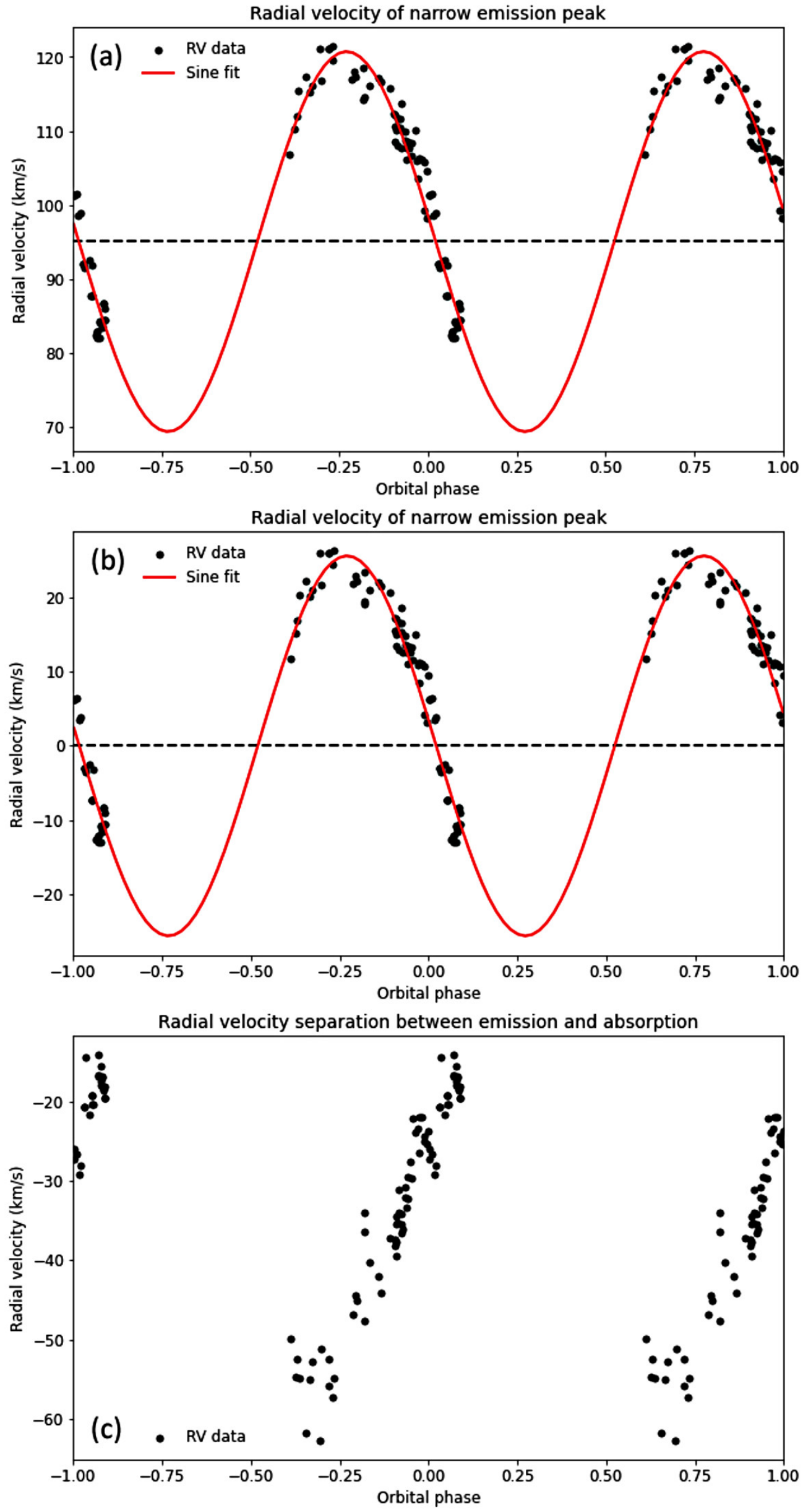

Figure 15. Sine fits to the radial velocities of the narrow Hα emission as measured (a) and in the system rest frame (b). Separation in radial velocity between emission and absorption in the system rest frame (c).

In the active phase following the probable outburst in 2025 there is an optically thin bipolar stellar wind from the hot component which follows it in orbit (Skopal 2006). UV radiation from the white dwarf ionizes this wind above and below the disk forming H II regions in which the low ionization energy recombination lines of H I and Fe II seen in Figure 6 are formed (Skopal 2003; Sokoloski 2003). The velocity dispersion of the wind generates narrow Hα emission with a FWHM of ~120 km/s reducing to ~85 km/s during the eclipse (Figure 16c). The radial velocity of this emission follows the white dwarf orbit around the systemic center of mass (Figure 16d).

The narrow Hα emission experiences blue-shifted P Cygni absorption. The separation in radial velocity between emission and absorption in the system rest frame is shown in Figure 15c and appears to follow a sinusoidal path. Around orbital phase 0.0 when the red giant is closer to us, the blue-shift is ~–20 km/s. Around phase 0.5 where the white dwarf is closer to us, the blue-shift is ~–60 km/s.

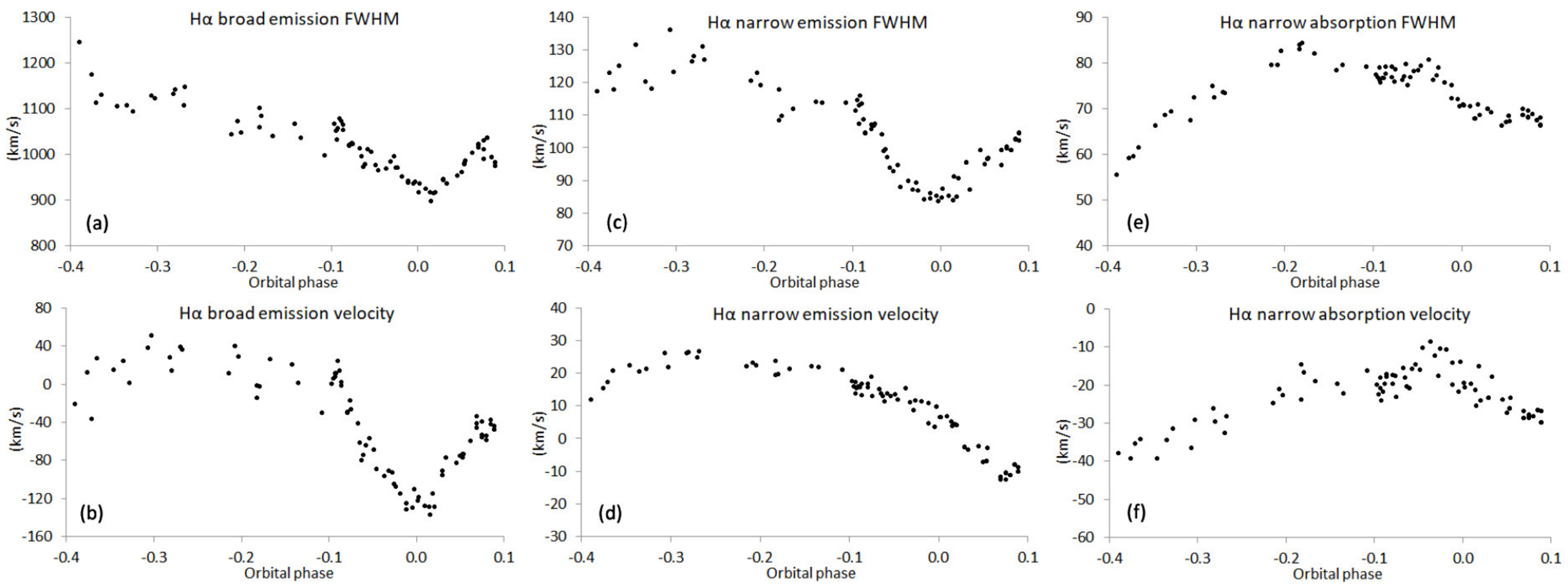

Figure 16. Variation with orbital phase of the FWHM and radial velocities of the broad emission (a, b), narrow emission (c, d), and narrow absorption (e, f) components of the Hα line in high-resolution spectra after subtraction of the systemic radial velocity.

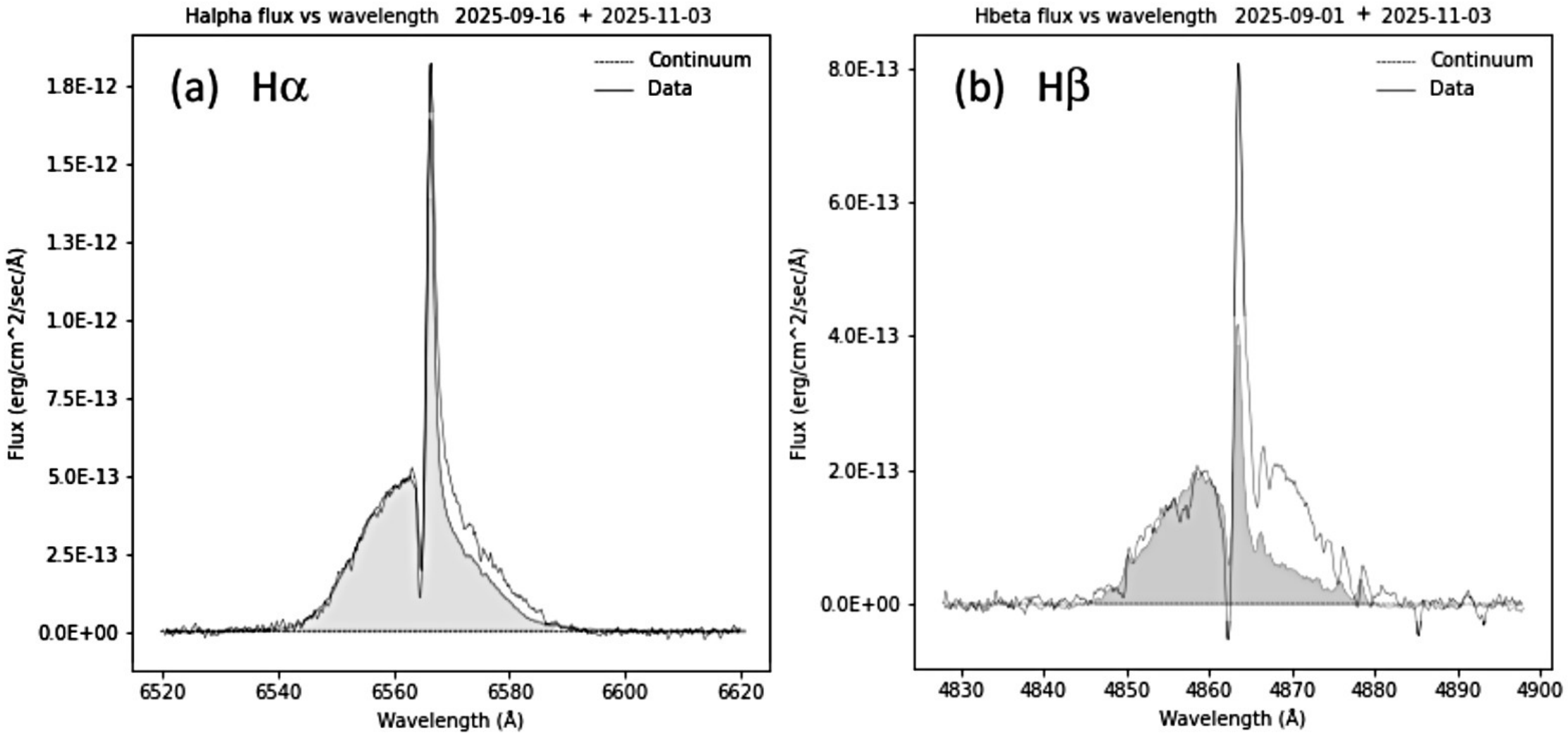

Figure 17. Profiles of the Hα and Hβ emission lines before and during the eclipse. During the eclipse (shaded) the long wavelength side of the line is depleted, causing the mean wavelength of the broad component of the line to move to shorter wavelengths and its radial velocity to become more negative.

The behavior of all three components of the Hα emission line can therefore be explained in terms of their formation in the environment around the expanded hot component following a probable symbiotic outburst early in 2025.

## 4. Summary

Using a combination of high and low resolution spectroscopy obtained by an international team of amateur astronomers, we recorded and analyzed 150 flux-calibrated spectra of the symbiotic binary V1413 Aql as it passed through an eclipse in November 2025 following a probable Z And-type symbiotic outburst. We measured this eclipse time of minimum, calculated the times of a further 16 previous eclipses from the AAVSO AID, and produced an updated eclipse ephemeris. We calibrated our spectra in absolute flux and used Gaussian fits to resolve and analyze components of the Hα emission line. Their observed behavior was consistent with following a Z And-type symbiotic outburst. We determined the radial velocity curve of the white dwarf measured the semi-amplitude as 25.7 km/s and calculated the systemic radial velocity as 95.1 km/s. The continuum of our low resolution spectra was consistent with that of a late F or early G type giant star.

## 5. Acknowledgements

The authors are grateful for helpful and informative comments from the anonymous referee.

We are also grateful for insightful comments from Professor Dr. Hans Martin Schmid and Dr. Chris Lloyd which have helped us to improve the paper. We acknowledge with thanks use of the ARAS Spectral Database, the BAA Spectroscopy Database and the AAVSO AVSpec Database. We also acknowledge with thanks the variable star observations from the AAVSO AID contributed by observers worldwide and used in this research.

## 6. Update of 2026 July 30

Further spectra of V1413 Aql obtained since the paper was submitted have enabled us to carry out a binary orbital analysis of the radial velocities of the narrow emission component using Binary Star Solver (Milson 2020) replacing the original sine fit in section 3.6. This gives revised values for the systemic velocity of 92.70 ± 0.29 km/s and for the semi-amplitude of 28.07 ± 0.45 km/s, and for the first time a value for the orbital eccentricity of 0.040 ± 0.012, confirming that the orbit of V1413 Aql is almost certainly circular.

## References

ARAS Spectral Database. 2026 (https://aras-database.github.io/database/index.html).

Astropy Collaboration, *et al.* 2018, *Astron. J.*, **156**, 123.

Bailer-Jones, C. A. L., Rybizki, J., Fouesneau, M., Demleitner, M., and Andrae, R. 2021, *Astron. J.*, **161**, 147.

Bisikalo, D. V., Boyarchuk, A. A., Kilpio, E. Yu., Tomov, N. A., and Tomova, M. T. 2006, *Astron. Rep.*, **50**, 722 (https://arxiv.org/abs/astro-ph/0610415).

Boyd, D. 2020, "A method of calibrating spectra in absolute flux using V magnitudes" (https://britastro.org/sites/default/files/absfluxcalibration.pdf).

Boyd, D., Cejudo, D., Foster, J., Sims, F., and Walker, G. 2025, *J. Amer. Assoc. Var. Star Obs.*, **53**, 31 (Paper 1).

British Astronomical Association. 2026, BAA Spectroscopy Database (https://britastro.org/specdb).

Cardelli, Jason A., Clayton, Geoffrey C., and Mathis, John S. 1989, *Astrophys. J.*, **345**, 245.

Eggleton, P. P. 1983, *Astrophys. J.*, **268**, 368.

Esipov, V. F., Kolotilov, E. A., Mikolajewska, J., Munari, U., Tatarnikova, A. A., Tatarnikov, A. M., Tomov, T., and Yudin, B. F. 2000, *Astron. Lett.*, **26**, 162.

Fekel, F. C., Hinkle, K. H., Joyce, R. R., and Skrutskie, M. F. 2000, *Astron. J.*, **120**, 3255.

Kloppenborg, B. 2026a, variable star observations from the AAVSO International Database (https://apps.aavso.org/v2/data/search/photometry/).

Kloppenborg, B. 2026b, AVSpec, variable star observations from the AAVSO Spectral Database (https://apps.aavso.org/avspec/search).

Merc, J. 2025, *Galaxies*, **13**, 49.

Mikołajewska, J. 2003, in *Symbiotic Stars Probing Stellar Evolution*, eds. R. L. M. Corradi, R. Mikolajewska, T. J. Mahoney, ASP Conf. Ser., 303, Astronomical Society of the Pacific, San Francisco, 9.

Milson, N. 2020, Binary Star Solver (https://github.com/NickMilsonPhysics/BinaryStarSolver).

Munari, U. 1992, *Astron. Astrophys.*, **257**, 163.

Poyner, G. 2023, *Br. Astron. Assoc Var. Star Section Circ.*, **197**, 23 (https://britastro.org/vss/VSSC197.pdf).

PyAstronomy. 2026, PyAstronomy (PyA), astronomy related packages hosted on github (https://pyastronomy.readthedocs.io/en/latest).

Skopal, A. 2003, "The role of ionization in symbiotic binaries" (https://arxiv.org/abs/astro-ph/0308462).

Skopal, A. 2006, *Astron. Astrophys.*, **457**, 1003.

Skopal, A. 2023, *Astron. J.*, **165**, 258.

Skopal, A., *et al.* 2011, *Astron. Astrophys.*, **536A**, 27.

Sokoloski, J. L. 2003, *J. Amer. Assoc. Var. Star Obs.*, **31**, 89.

Sokoloski, J. L., *et al.* 2006, *Astrophys. J.*, **636**, 1002.

Spectroscopy Discussion Group. 2026, Spectroscopy Discussion Group (https://aatel.groups.io/g/SpectroscopyDiscussionGroup).

Tatarnikova, A., Tatarnikov, A., Maslennikova, N., Dodin, A., Burlak, M., Ikonnikova, N., Belinski, A., and Nikishev, G. 2025, *Galaxies*, **13**, 86.

Teyssier, F. 2019, *Contrib. Astron. Obs. Skalnaté Pleso*, **49**, 217.

van Belle, G. T., *et al.* 1999, *Astron. J.*, **117**, 521.

Zamanov, R. K., Bode, M. F., Melo, C. H. F., Bachev, R., Gomboc, A., Stateva, I. K., Porter, J. M., and Pritchard, J. 2007, *Mon. Not. Roy. Astron. Soc.*, **380**, 1053.